\documentclass[a4paper,aps,showpacs,superscriptaddress,floatfix,nofootinbib,twocolumn,bibtex]{revtex4}

\usepackage{bbold}
\usepackage{bbm}
\usepackage[pdftex]{graphicx}
\usepackage{latexsym,amsmath,verbatim}
\usepackage{color}
\usepackage{rotating}
\usepackage{multirow}
\usepackage[english]{babel}

\begin{document}

\title{Detectability of communities in heterogeneous networks}

\author{Filippo Radicchi}\email{f.radicchi@gmail.com}
\affiliation{Departament d'Enginyeria Quimica, Universitat Rovira i Virgili, 43007 Tarragona, Spain}

\date{\today}

\begin{abstract}
\noindent 
Communities are fundamental entities for the characterization
of the structure of real networks.
The standard approach to the identification of communities 
in networks is based on the optimization of a quality function 
known as ``modularity''.
Although modularity has been at the center
of an intense research activity and many methods
for its maximization have been proposed, 
not much it is yet known about the necessary
conditions that communities need to satisfy in order
to be detectable with modularity maximization methods.
Here, we develop a simple
theory to establish 
these conditions,
and we successfully apply it to
various classes of network models. 
Our main result is that heterogeneity in the
degree distribution helps modularity
to correctly recover the community structure
of a network and that, in the realistic 
case of scale-free networks with degree exponent $\gamma < 2.5$,
modularity is always able to detect the presence of communities.     
\end{abstract}


\maketitle

\noindent  Communities are organizational modules that provide a 
coarse grained view of a complex 
network~\cite{Girvan02, Radicchi04, Fortunato10}.
Depending on the nature of the network, communities can
have different yet fundamental meanings:
in 
biological networks, communities are
likely to group 
entities having the same 
biological
function~\cite{Spirin03, Huberman04, Guimera05}, in the graph of the World
Wide Web they may correspond to groups of pages dealing
with the same or related topics~\cite{Dourisboure09}, 
in food webs they may identify compartments~\cite{Stouffer12}, etc.
Since communities play an important role for the characterization
of the structure of networks, 
the development of
computer algorithms for the detection
of communities in networks represents
one of the most active areas
in network science~\cite{Fortunato10}. 
In particular, 
the use  of the so-called ``modularity'' 
has attracted a great attention
in recent years~\cite{Newman04, Newman06}. 
Modularity is a quality
function that estimates the relevance of
a given network partition (i.e., a division of the 
network in a given set of communities) by comparing the observed number
of internal connections 
of the communities 
with the expected number of such edges
in a random annealed version of the network. 
The best community division of the network is then given
by the partition that maximizes the modularity function.
\\
Although subjected to 
some intrinsic
limitations~\cite{Fortunato07,Clauset10}, modularity has become a standard tool
for community detection and several methods
for modularity maximization
have been developed~\cite{Fortunato10}.
In this paper, we focus our attention on spectral
optimization of the modularity function, i.e.,  
a maximization method that
is essentially based on the determination
of the principal eigenpair of the so-called modularity matrix~\cite{Newman06}. 
This represents a way to approximate the configuration
corresponding to the maximum of the
modularity function, whose determination
would be otherwise a NP complete problem~\cite{Brandes08},
by relaxing the indices that assign the nodes
to the various communities from integer
to real valued numbers. This method provides
in general solutions that are consistent
with those obtained by other more sophisticated
maximization techniques~\cite{Newman06, Newman12}, thus
the following results can be
reasonably considered as valid for
any type of community detection algorithm based on
modularity maximization.

\

\noindent We consider here the
simple case of a
symmetric and weighted network formed only by
$2$ communities of size $N$.
The adjacency matrices that contain the information
about the internal structure 
of these two groups are denoted
respectively with $A_{1,1}$ and $A_{2,2}$, while
the connections among nodes of different groups
are listed in the matrix $A_{1,2}= A_{2,1}^T$.
The adjacency matrix of the entire network can
be thus written in the following block form
\begin{equation}
A  = \left(
\begin{array}{cc}
A_{1,1}  & A_{1,2}\\
A_{2,1} & A_{2,2}
\end{array}
\right) \; .
\label{eq:adj}
\end{equation}

\noindent According to the definition of the modularity
function, in the random annealed version of the network,
which preserves on average the node degrees, the probability
that two nodes are connected is proportional
to the product of the degrees of the two nodes~\cite{Molloy95, Newman04}.
The entire information of this null model
is contained in the square matrix 
\[
P = \frac{1}{\left<s_1|1\right> + \left<s_2|1\right>}\left(
\begin{array}{cc}
\left|s_1\right>\left<s_1\right| & \left|s_1\right>\left<s_2\right| \\
\left|s_2\right>\left<s_1\right| & \left|s_2\right>\left<s_2\right|
\end{array}
\right) 
\; ,
\]
where $\left|s_i\right> = \left|s_{i,i}\right> + \left|s_{i,j}\right>$
is the strength vector of the $i$-th group, with
$\left|s_{i,j}\right> = A_{i,j} \left|1\right>$
equal to a vector whose components 
are equal to the sum of the weights 
of all edges
connecting nodes of the $i$-th group to nodes
of the $j$-th group, and
$\left|1\right>$ is
the vector whose components are all equal to one.

\

\noindent Let us focus our attention on the spectrum
of the modularity matrix $Q=A-P$~\cite{Newman06}. 
Note that by definition $Q \left|1\right> = \left|0\right>$,
where $\left|0\right>$ is the vector with all
components equal to zero,  
thus any other eigenvector $\left|v\right>$
of the modularity matrix $Q$, i.e.,
$Q \left|v\right> = \lambda \left|v\right>$,
must be orthogonal to the vector $\left|1\right>$.
We can rewrite the eigenvector 
$\left|v\right> = \left|v_1, v_2\right>$,
where $\left|v_i\right>$ is 
the part of the eigenvector
$\left|v\right>$ that corresponds to the $i$-th group.
The orthogonality with respect to
the eigenvector $\left|1\right>$ reads
$\left<v_1|1\right> + \left<v_2|1\right> = 0$, 
while the normality of the eigenvector means
that
$\left<v_1|v_1\right> + \left<v_2|v_2\right>  = 1$. 
The eigenvalue problem becomes equivalent to
\begin{equation}
A_{1,1} \left|v_1\right> + A_{1,2} \left|v_2\right> - \frac{\left<s_1|v_1\right> + \left<s_2|v_2\right>}{\left<s_1|1\right> + \left<s_2|1\right>} \left|s_1\right> = \lambda \left|v_1\right>
\label{eq:eig1}
\end{equation}
and
\begin{equation}
A_{2,2} \left|v_2\right> + A_{2,1} \left|v_1\right> - \frac{\left<s_1|v_1\right>+ \left<s_2|v_2\right>}{\left<s_1|1\right> + \left<s_2|1\right>} \left|s_2\right> = \lambda \left|v_2\right> \; .
\label{eq:eig2}
\end{equation}
If we multiply them for $\left<1\right|$, and then 
take their difference, we obtain

\begin{equation}
\begin{array}{l}
\lambda \left(\left<v_1|1\right> -\left<v_2|1\right>\right) =   \left( \left<s_{1,1}|v_1\right> - \left<s_{1,2}|v_1\right>  \right.  
\\
\left. - \left<s_{2,2}|v_2\right> + \left<s_{2,1}|v_2\right>\right) - \alpha 
\left(  \left<s_{1}|v_1\right> - \left<s_{2}|v_2\right>  \right)
\end{array} \; ,
\label{eq:diff}
\end{equation}
where we have defined $\alpha = \frac{\left<s_1|1\right> - \left<s_2|1\right>}{\left<s_1|1\right> + \left<s_2|1\right>}$.
For simplicity, in the following we will
consider only cases in which $\alpha \simeq 0$, i.e.,
cases in which both groups have a comparable total number of edges.

\

\begin{figure}[!htb]
\includegraphics[width=0.45\textwidth]{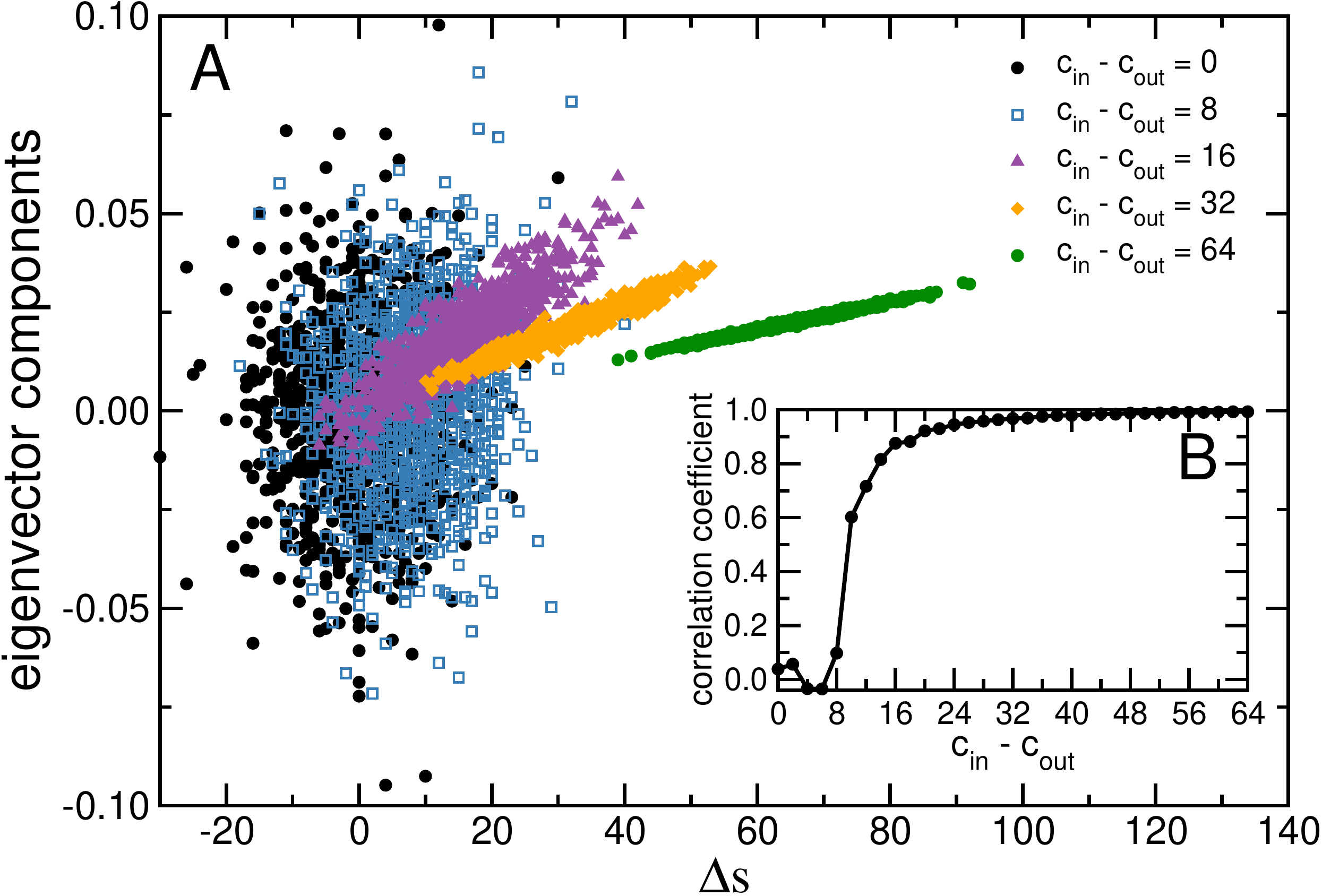}
\caption{(Color online) {\bf A} Components of the principal
eigenvector of the modularity matrix as functions of the difference between
internal and external degrees for a system 
composed of two poissonian networks of
size $N=1024$, with average internal degree $c_{in}$
and average external degree $c_{out}$ such that
$c_{in}+c_{out}=64$. We plot here only the components of one of the
two modules, but analogous
results (with opposite sign) 
are valid for the other module.
Different colors and symbols correspond to
different combinations 
of $c_{in}$ and $c_{out}$.
{\bf B} Correlation coefficient between
the eigenvector components and the
difference between internal and external degrees 
as a function of  
$c_{in} - c_{out}$. While the correlation coefficient
is not significant in the region
in which communities are not detectable [i.e., $c_{in}-c_{out} < 
\sqrt{c_{in}+c_{out}}$, see Eq.~(\ref{eq:poisson_t})],
it becomes significant in the detectability regime 
(i.e., $c_{in}-c_{out} \geq 
\sqrt{c_{in}+c_{out}}$).}
\label{fig1}
\end{figure}

\noindent We define the detectability
regime as the regime in which the two 
pre-imposed communities
can be detected by means of modularity 
spectral optimization.
This regime is characterized by the fact that
the components of the principal eigenvector corresponding
to the nodes of one of the modules 
have coherent signs, while the 
two portions of the eigenvector
corresponding to different groups are opposite
in sign~\cite{Newman06}.
If we suppose that
these modules are uncorrelated graphs (i.e.,
without further internal sub-community structure) with 
prescribed in- and out-strength vectors,
we expect that this eigenvector 
is such that
\begin{equation}
\left|v_1\right> =  n_1 \, \left( \left|s_{1,1}\right> - \left|s_{1,2}\right> \right)  = n_1 \left|\Delta s_1\right>  
\label{eq:prop1}
\end{equation}
and 
\begin{equation}
 \left|v_2\right>  = n_2 \,   \left( \left|s_{2,2}\right> - \left|s_{2,1}\right> \right)  = n_2 \left|\Delta s_2\right>  \;, 
\label{eq:prop2}
\end{equation}
where $n_1$ and $n_2$ are proportionality constants, while
$\left|\Delta s_1\right>$ and $\left|\Delta s_2\right>$
are respectively the vectors whose entries
are given by the difference of the in- and out-strengths
of the nodes in the groups 1 and 2.
Eqs.~(\ref{eq:prop1}) and~(\ref{eq:prop2}) 
simply state that the coordinates of
$\left|v_1\right>$ and $\left|v_2\right>$ are
linearly proportional to the difference of the in- and out-
strength vectors, a solution that appears
natural if we interpret
the modularity function as the stationary 
solution of a random walk between
the two communities~\cite{Lambiotte09, Lambiotte10, Mucha2010}.
This conjecture is indeed perfectly verified 
in numerical estimations of the largest eigenvector of the
modularity matrix (see Figs.~\ref{fig1},S1 and S2),
and thus 
Eqs.~(\ref{eq:prop1}) and~(\ref{eq:prop2})  
can be used as a reasonable ansazt
for the solution of our problem.
If we finally insert Eqs.~(\ref{eq:prop1}) and~(\ref{eq:prop2})
into Eq.~(\ref{eq:diff}), we
can give an estimate of the largest 
eigenvalue $\lambda_{max}$ of the modularity matrix
in the regime of detectable communities, and write
\begin{equation}
\lambda_{max} = \frac{1}{2} \left(\frac{ \left<\Delta s_{1}|\Delta s_{1}\right>}{ \left<\Delta s_{1}|1\right>}  +  \frac{ \left<\Delta s_{2}|\Delta s_{2}\right>  } {\left<\Delta s_{2}|1\right>} \right)\; ,
\label{eq:approx}
\end{equation} 
where we have used the orthogonality 
condition $\left<v_1|1\right> + \left<v_2|1\right>=0$, which leads to
$n_1 = - n_2 \left<\Delta s_2|1\right> / \left<\Delta s_1|1\right>$.
Note that expression Eq.~(\ref{eq:approx}) is valid
for given strength vectors. If we instead assume
that the entries of these vectors
are random variates obeying the statistical 
distributions $P\left(\Delta s_1\right)$
and $P\left(\Delta s_2\right)$, we can write
\begin{equation}
\lambda_{max} = \frac{1}{2} \left(\frac{  m_2^{(1)}}{m_1^{(1)} }  + \frac{  m_2^{(2)} } {m_1^{(2)}} \right)\; ,
\label{eq:ensemble}
\end{equation} 
where $m_1^{(i)}$ and $m_2^{(i)}$ are respectively the first 
and the second moments of
the distribution $P\left(\Delta s_i\right)$.

\

\noindent It is important to stress that Eqs.~(\ref{eq:approx})
and~(\ref{eq:ensemble}) give us an estimate of
the largest eigenvalue of $Q$ only in the detectability regime.
If the structure of the entire graph is instead such
that the two modules are not detectable by means
of modularity maximization, there will another
principal eigenvector orthogonal to the previous one, and
thus not showing the presence of the two modules.
Since we have supposed that both modules
are randomly generated graphs, the other
eigenvalue that is competing with $\lambda_{max}$
for being the highest eigenvalue of the modularity matrix
is given by the second largest eigenvalue
of the annealed random network associated to $Q$~\cite{Newman12}.
In intuitive terms, this means that, in the regime in which
the groups are  undetectable,
the signal
present in $A$ is not sufficiently high, and $Q$ is in spectral
terms
indistinguishable from $P$.
In the following, we will
consider some examples of network ensembles where 
both these eigenvalues can be analytically estimated, and thus
the detectability problem can be explicitly solved. 

\

\begin{figure}[!htb]
\includegraphics[width=0.45\textwidth]{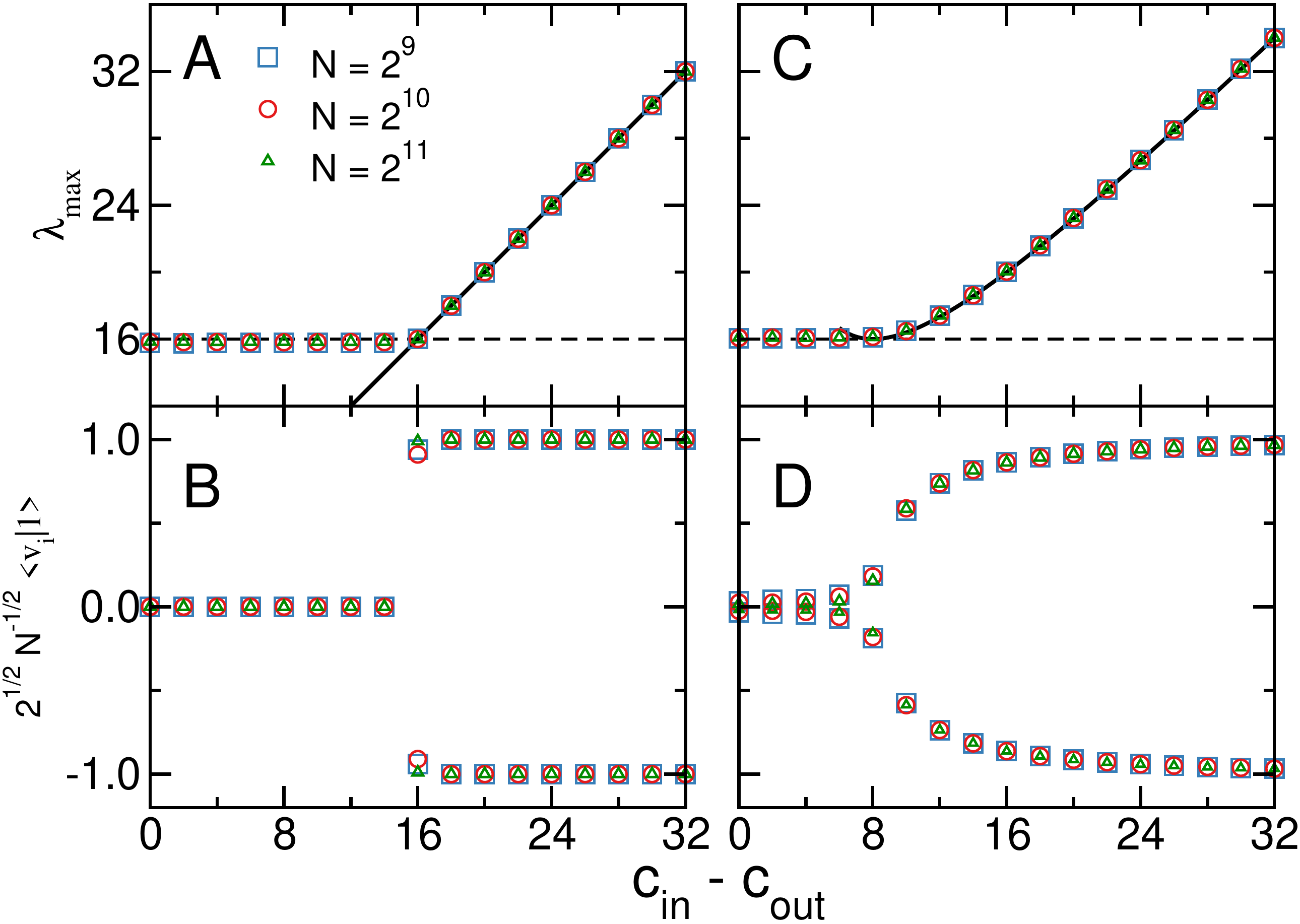}
\caption{(Color online) Numerical estimation of the 
largest eigenpair of the modularity matrix $Q$
in the case of two random regular graphs (panels A and B)
and in the case of two poissonian graphs (panels C and D).
In both cases we consider three different network sizes
$N=256$ (blue squares), $N=512$ (red
circles) and $N=1024$ (green triangles),
and we fix $c_{in}+c_{out}=64$. The results
presented here
have been obtained by averaging over $100$ different
realizations of the models.
{\bf A} and {\bf C} Largest
eigenvalue of $Q$ as a function of $c_{in}-c_{out}$.
The full black line corresponds to $c_{in}-c_{out}$ in panel A, 
and to Eq.~(\ref{eq:poisson}) 
in C. The dashed black lines are given by $2\sqrt{c_{in}+c_{out}}$.
{\bf B} and {\bf D} Size independent inner products
$\left<v_1|1\right>$ and $\left<v_2|1\right>$  
as  functions
of $c_{in}-c_{out}$. 
}
\label{fig2}
\end{figure}

\noindent {\it Regular graphs}. In this case,  
each node has exactly $c_{in}$ random connections
with other nodes in its group, and
$c_{out}$ random connections outside its own group.
Eq.~(\ref{eq:diff}) reduces to 
\begin{equation}
\begin{array}{l}
\left(\lambda_{max} - c_{in} + c_{out}\right) \left<v_1|1\right> = 0
\\
\left(\lambda_{max} - c_{in} + c_{out}\right) \left<v_2|1\right> = 0 
\end{array}
\;,
\label{eq:regular}
\end{equation}
thus either (i) $\left<v_1|1\right>=\left<v_2|1\right>=0$ and $\lambda_{max} \neq c_{in}-c_{out}$, or (ii) $\lambda_{max}= c_{in}-c_{out}$, $\left<v_1|1\right> \neq 0$ and $\left<v_2|1\right> \neq 0$. In case (ii), one can also prove
that the only possible solution  
is 
$
\left|v_1\right> = \pm \left|1\right>/ \sqrt{2N}$
and $\left|v_2\right> = \mp \left|1\right> / \sqrt{2N}$ (see Supplemental Material).
The same result can be also obtained using 
Eq.~(\ref{eq:ensemble}) that reduces 
to Eq.~(\ref{eq:regular}) by setting
$P\left(\Delta s_1\right) = P\left(\Delta s_2\right) = 
\delta\left(c_{in} - c_{out}\right)$.
The term of comparison for $\lambda_{max}$ in the case
of regular graphs is given by the second largest
eigenvalue of the adjacency matrix of a random regular
graph with valency $c=c_{in} + c_{out}$, that is 
in good approximation equal to
$2 \sqrt{c}$~\cite{alon86, Friedman03aproof}.
Eq.~(\ref{eq:regular}) tells us that,
independently of the system size, the two modules can be either fully
detectable or not detectable at all. The sudden transition
between these two regimes happens at the point in which
\begin{equation}
c_{in}-c_{out} = 2 \sqrt{c_{in}+c_{out}} \; .
\label{eq:regular_t}
\end{equation}
This theoretical prediction is in perfect agreement with the results
of the numerical simulations reported in Fig.~\ref{fig2}.

\

\noindent {\it Poissonian networks}. This is 
the case whose entire spectrum has been
analytically determined by Nadakuditi and Newman~\cite{Newman12}.
Internal degrees in both groups are drawn from a Poisson distribution
with average $c_{in}$, and external degrees 
are also poissonian variates with
average $c_{out}$. In this case, the distributions 
$P\left(\Delta s_1\right)$
and $P\left(\Delta s_2\right)$ are two identical
Skellam distributions, with first
moment equal to $m_1 = c_{in} - c_{out}$, 
and second moment equal to
$m_2 = \left(c_{in}+c_{out}\right) + \left(c_{in}-c_{out}\right)^2$~\cite{Skellam46}.
We can 
reduce Eq.~(\ref{eq:ensemble}) to
\begin{equation}
\lambda_{max} = c_{in} - c_{out} +   \frac{ c_{in}+c_{out}  } {c_{in} - c_{out}} \; .
\label{eq:poisson}
\end{equation} 
This result is identical to the
prediction obtained in~\cite{Newman12}.
The term of comparison for the largest eigenvalue of the modularity
matrix is given by the second largest eigenvalue
of a random graph with average degree $c=c_{in}+c_{out}$, 
that is $2\sqrt{c}$~\cite{Chung03, Newman12}, and this
finally leads to the detectability threshold
\begin{equation}
c_{in} - c_{out} = \sqrt{c_{in}+c_{out}} \; ,
\label{eq:poisson_t}
\end{equation} 
as already obtained in~\cite{Decelle11, Newman12}.
The results of numerical simulations
perfectly agree with our theoretical prediction 
(see Fig.~\ref{fig2}). The prediction appears
to be not visibly dependent on the system
size, and already for small networks
Eq.~(\ref{eq:poisson_t}) represents a very
good estimate of the transition point.
We note that, as in the case of regular graphs, 
for a large portion of the region $c_{in}>c_{out}$ modularity
fails to recover the community structure of the graph.
It is, however, interesting to stress
that the detectability  threshold is two times
smaller than the one registered for regular graphs, 
and thus the heterogeneity in node degrees seems to enhance 
the ability of modularity to detect communities.

\

\begin{figure}[!htb]
\includegraphics[width=0.45\textwidth]{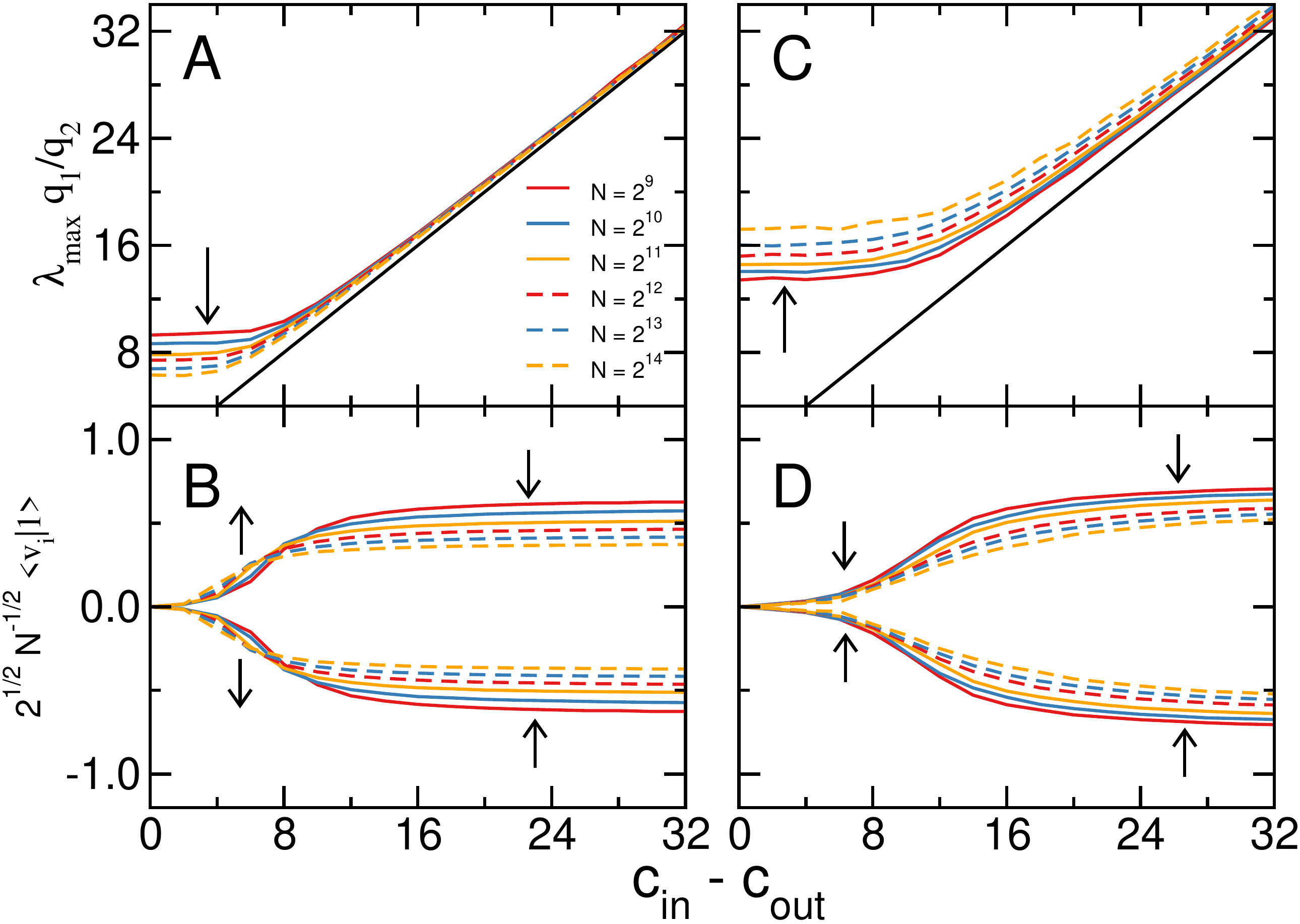}
\caption{(Color online) Numerical estimation of the 
largest eigenpair of the modularity matrix $Q$
in the case of two random scale-free graphs 
with degree exponents $\gamma=2.2$ (panels A and B)
and $\gamma=2.8$ (panels C and D).
In both cases, we consider several network sizes
ranging from $N=256$ to $N=8192$, and we set
$c_{in}+c_{out}=64$. The results presented here
have been obtained  by averaging over $100$ different
realizations of the models. To suppress
fluctuations, we restricted to the
case of networks for which 
the square of the largest strength
is smaller than the sum of all strengths 
(i.e., $s_{max}^2 < \sum_i s_i$)~\cite{Chung03, Chung03a},
although the qualitative outcome does not depend on this choice. 
For clarity, we placed arrows 
in the various panels to indicate
the direction of increasing $N$.
{\bf A} and {\bf C} Largest
eigenvalue $\lambda_{max}$ multiplied by $q_1/q_2$ 
as a function of $c_{in}-c_{out}$.
The full black line corresponds to $c_{in}-c_{out}$.
{\bf B} and {\bf D} Size independent inner products
$\left<v_1|1\right>$ and $\left<v_2|1\right>$  
as  functions
of $c_{in}-c_{out}$. 
}
\label{fig3}
\end{figure}

\noindent {\it LFR benchmark graphs}. As a final example, 
we consider a special case of the benchmark graphs
introduced by Lancichinetti {\it et al}~\cite{Lancichinetti08}. We set
$\left|s_{1,1}\right> = \left|s_{2,2}\right> = c_{in} \left|s\right>$
and $\left|s_{1,2}\right> = \left|s_{2,1}\right> = c_{out} \left|s\right>$,
where the entries of the vector $\left|s\right>$ are
random variates in the range $1$ to $N$ 
taken from a power-law distribution with exponent $\gamma$.
Eq.~(\ref{eq:ensemble}) becomes
\begin{equation}
\lambda_{max} = \left(c_{in}-c_{out}\right) \, \frac{\zeta_N\left(\gamma-2\right)}{\zeta_N\left(\gamma-1\right)} \; ,
\label{eq:lfr}
\end{equation} 
where $\zeta_N\left(x\right) = \sum_{n=1}^{N} n^{-x}$ 
is the Riemann zeta function truncated at the $N$-th term.
The term of comparison for the largest eigenvalue of the modularity
matrix is still given by the second largest eigenvalue
of the annealed random graph
associated with the modularity matrix.
We do not have an exact guess on how this
quantity depends on parameters of the network model, but
we can use the upper bound of the
largest eigenvalue of random scale-free graphs
to get more insights.
According to the predictions by
Chung {\it et al} adapted to the
present case, the
largest eigenvalue $\mu_{max}$ of our
random scale-free graphs is equal to
the maximum between
$\mu_{max}^{(1)}=\left(c_{in}+c_{out}\right) \, \frac{\zeta_N\left(\gamma-2\right)}{\zeta_N\left(\gamma-1\right)}$
and $\mu_{max}^{(2)}= \sqrt{\left(c_{in}+c_{out}\right) s_{max}}$,
with $s_{max}$ largest degree
in the network~\cite{Chung03, Chung03a}. In the
limit of sufficiently large $N$, we have
that: for $\gamma < 2.5$, the dominating 
eigenvalue is $\mu_{max}^{(1)}$; for $\gamma>2.5$, 
the largest
eigenvalue is instead $\mu_{max}^{(2)}$.
This has very important implications
when compared to our prediction of $\lambda_{max}$
given in Eq.~(\ref{eq:lfr}): 
(i) For $\gamma < 2.5$, $\lambda_{max}$ grows
as fast as $\mu_{max}$ with the system size,
thus the detectability threshold should approach zero
as $N$ increases. (ii) For $\gamma > 2.5$ instead, $\mu_{max}$
grows faster than $\lambda_{max}$ as $N$ increases.
The detectability threshold should grow
with the system size, and eventually converge 
to a finite fixed value (for instance, for $\gamma \to \infty$
we must recover the result valid for
the case of regular graphs).
\\ 
The results of numerical simulations support
our thesis (see Fig.~\ref{fig3}). When we plot
$\lambda_{max} \frac{q_1}{q_2}$ as a function
of $c_{in}-c_{out}$, with $q_1$ and $q_2$ respectively
the first and second moments of the strength distribution
of the network, we see that when $\gamma<2.5$ this quantity
slowly approaches, as $N$ increases, 
the linear behavior 
$c_{in}-c_{out}$ as predicted by Eq.~(\ref{eq:lfr}). This means that, in the
limit of infinite large systems, modularity
is able to detect the presence of the
network blocks for every $c_{in}>c_{out}$.
Instead, for $\gamma>2.5$,
the lower part of the curve tends to move away
from the linear behavior as
the system size grows. This implies that
there will be, also in the limit of
infinitely large systems, always a part
of the $c_{in}>c_{out}$ region in which the two blocks
are undetectable via modularity maximization.

\

\noindent To summarize, we identified the
necessary conditions that communities
need to satisfy in order to be detectable by
means of modularity maximization. Our results
are valid for the case of $2$ groups with
comparable number of edges, and when
the information about the number of such groups
is used as ingredient in the maximization of the
modularity function.
Our main result is that in random network ensembles with 
pre-imposed community structure,
the eigenvector of the modularity matrix that identifies the
presence of the block structure is associated
with an eigenvalue approximately 
equal to the ratio between the
second and the first moments of the 
distribution of the difference between 
internal and external node strengths.
If this eigenvalue is larger than the second largest
eigenvalue of the null model associated to
the modularity function, then modularity
is able to detect such a structure, otherwise not.
This represents a limitation in the case
of graphs with homogeneous degrees. 
Increasing the heterogeneity of the network
accelerates instead the ability of
modularity to recover the correct 
community structure. For example,
adding noise to a regular graph makes
the detectability threshold two times smaller.
More importantly, 
if the heterogeneity of the 
node degrees is sufficiently
high, as in the case of real networked systems,
then modularity is always able to detect communities.

\begin{acknowledgements}
\noindent The author thanks A. Arenas, R. Darst and S. Fortunato for helpful
discussions on the subject of this article. The author acknowledges
support from the Spanish Ministerio de Ciencia e Innovaci\'on
through the Ram\'on y Cajal program.
\end{acknowledgements}


\newpage


\onecolumngrid

\section*{Supplemental Material}

\renewcommand{\theequation}{S\arabic{equation}}
\setcounter{equation}{0}
\renewcommand{\thefigure}{S\arabic{figure}}
\setcounter{figure}{0}
\renewcommand{\thetable}{S\arabic{table}}
\setcounter{table}{0}

\subsection*{Principal eigenpair of the modularity matrix for regular graphs}
\noindent For regular graphs, we
have
\[
\left|s_1\right> = \left|s_2\right>  = (c_{in}- c_{out}) \left|1\right> \;,
\]
thus 
\[
\left<s_1|v_1\right> + \left<s_2|v_2\right> = (c_{in}- c_{out}) \left( \left<1|v_1\right> + \left<1|v_2\right> \right) = 0 
\]
for the orthogonality of the eigenvector $\left|v_1,v_2\right>$ 
with respect to the
vector $\left|1\right>$, 
and Eqs.(2) and (3) of the main text
reduce to
\begin{equation}
A_{1,1} \left|v_1\right> + A_{1,2} \left|v_2\right> = \lambda \left|v_1\right>
\label{eq1}
\end{equation}
and 
\begin{equation}
A_{2,2} \left|v_2\right> + A_{2,1} \left|v_1\right> = \lambda \left|v_2\right> \; .
\label{eq2}
\end{equation}

\noindent 
Consider the eigenvalue $\lambda = c_{in} - c_{out}$ 
(which corresponds to the principal eigenvalue of the modularity
matrix in the case of detectable communities, as proved in the main text). 
The only term on the l.h.s. of 
Eq.~(\ref{eq1}) that depends on $c_{in}$ is $A_{1,1} \left|v_1\right>$, while
the only term that depends on $c_{out}$ is $A_{1,2} \left|v_2\right>$. This
means that Eq.~(\ref{eq1}) can be decoupled in two equations
\begin{equation}
A_{1,1} \left|v_1\right> = c_{in} \left|v_1\right> \qquad \textrm{ and } \qquad  A_{1,2} \left|v_2\right> = - c_{out} \left|v_1\right> \; .
\label{eq1a}
\end{equation}
Similarly, Eq.~(\ref{eq2}) leads to
\begin{equation}
A_{2,2} \left|v_2\right> = c_{in} \left|v_2\right> \qquad \textrm{ and } \qquad  A_{2,1} \left|v_1\right> = - c_{out} \left|v_2\right> \; .
\label{eq2a}
\end{equation}
Since the subgraphs encoded by the adjacency matrices $A_{1,1}$ and $A_{2,2}$
are regular graphs with valency $c_{in}$, this means that $\left|v_1\right> = n_1 \left|1\right>$
and $\left|v_2\right> = n_2 \left|1\right>$, with $n_1$ and $n_2$
suitable normalization constants. 
The same consideration is valid also for the subgraph encoded
by the adjacency matrix $A_{1,2} = A_{2,1}$ which is still a regular graph
(with valency $c_{out}$ in this case) and thus 
Eq.~(\ref{eq1a}) and (\ref{eq2a}) consistently lead to the same 
solutions $\left|v_1\right> = n_1 \left|1\right>$
and $\left|v_2\right> = n_2 \left|1\right>$.
Since the eigenvector $\left|v_1, v_2\right>$
is orthogonal to the vector $\left|1\right>$ (i.e, $\left<v_1|1 \right> + \left<v_2|1 \right> = 0$) and properly normalized
(i.e., $\left<v_1|v_1\right> + \left<v_2|v_2\right> =1$), one finally finds
that $n_1 = - n_2 = \pm 1/\sqrt{2N}$.

\newpage

\subsection*{Numerical estimation of the principal eigenpair of the modularity matrix for LFR benchmark graphs}

\begin{figure*}[!htb]
\includegraphics[height=0.28\textheight]{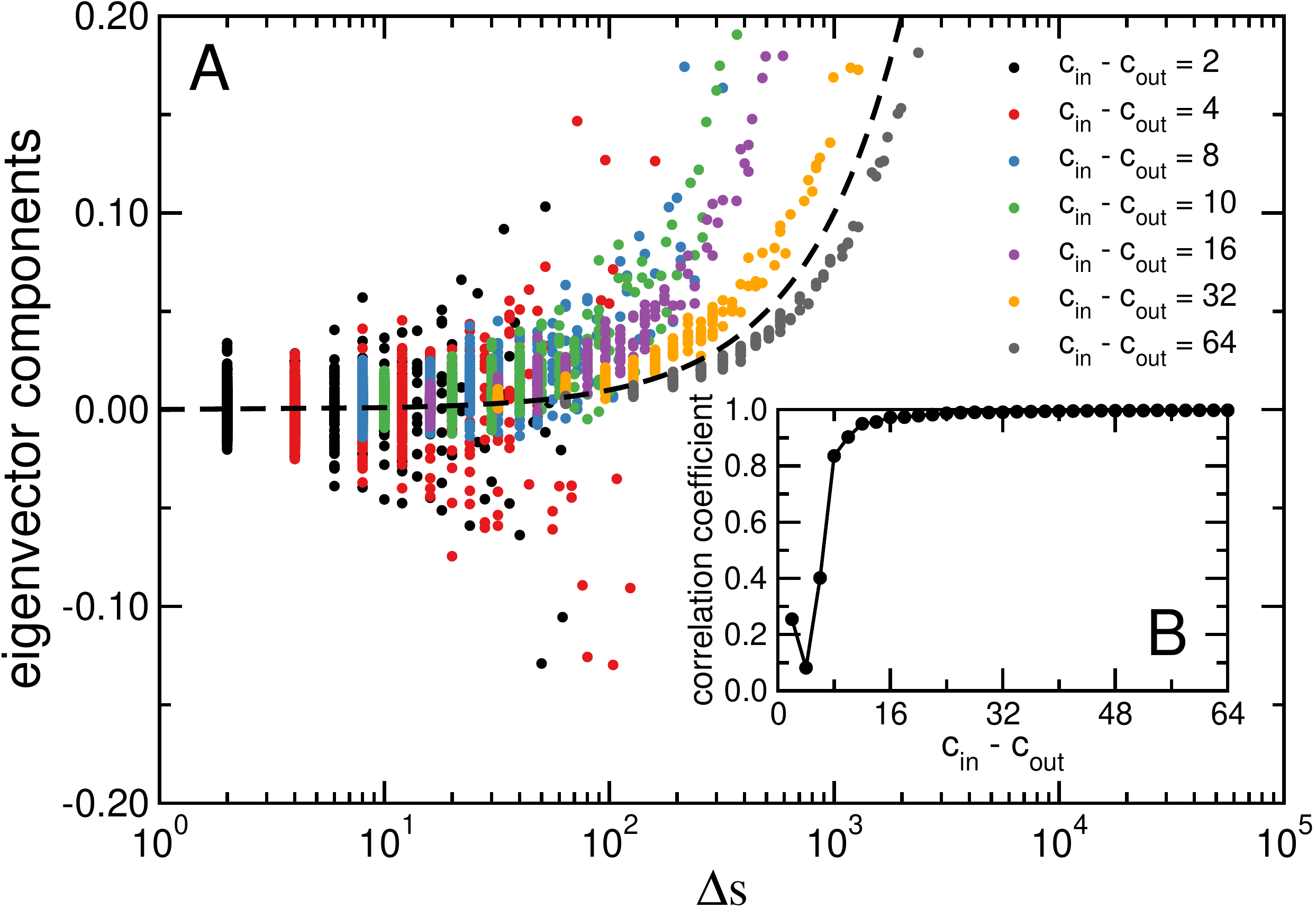}
\qquad
\includegraphics[height=0.28\textheight]{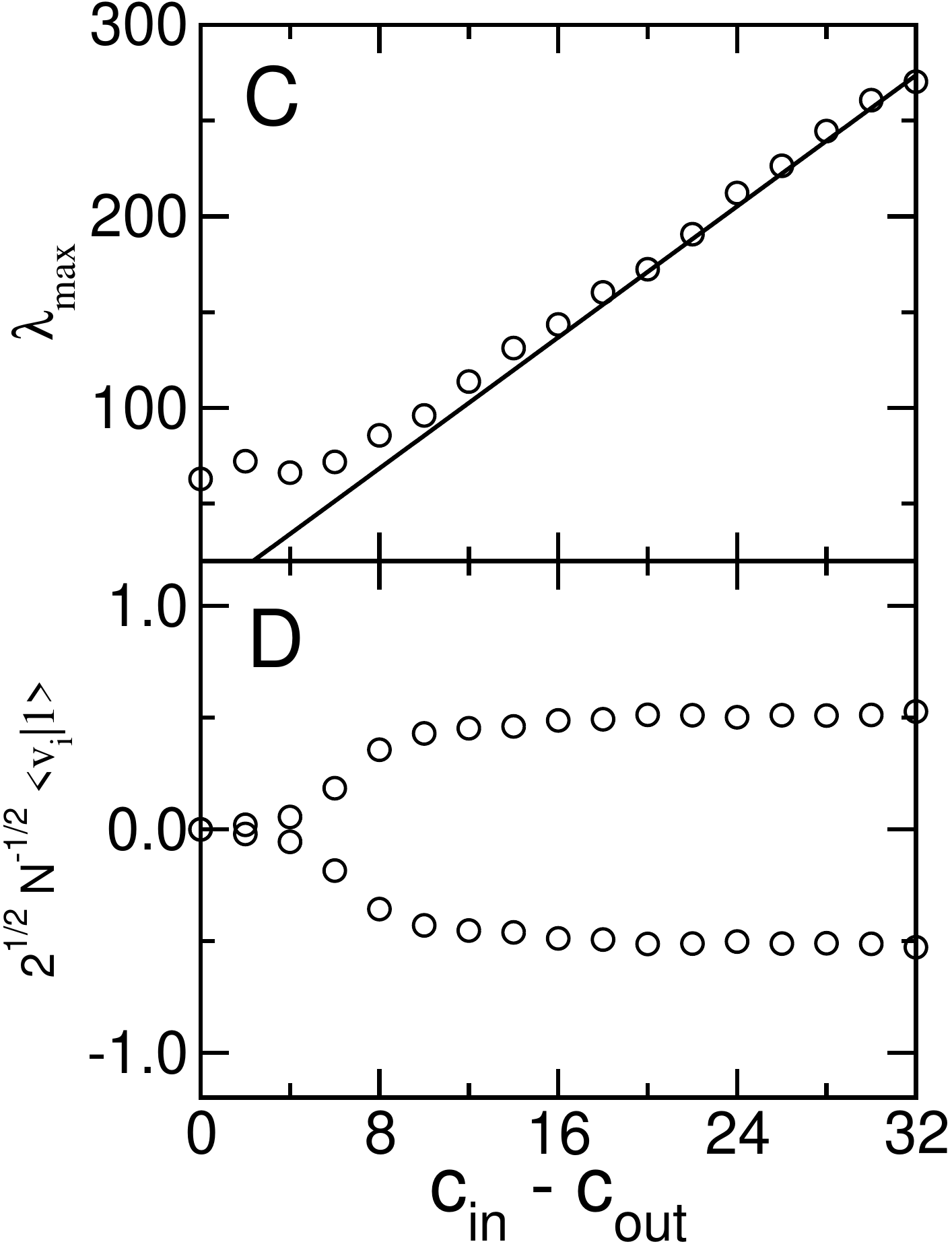}
\caption{Numerical estimation of the 
largest eigenpair of the modularity matrix $Q$
for a single realization of two random scale-free graphs 
with degree exponent $\gamma=2.2$, size $N=1024$,
and internal and external degree parameters such that $c_{in}+c_{out}=64$. 
{\bf A} Components of the principal
eigenvector as functions of the difference between
internal and external degrees. 
We plot here only the components of one of the
two modules, but analogous
results (with opposite sign) 
are valid for the other module.
Different colors corresponds to
different combinations 
of $c_{in}$ and $c_{out}$.  The black dashed line
is proportional to $c_{in}-c_{out}$.
{\bf B} Correlation coefficient between
the eigenvector components and the
difference between internal and external degrees 
as a function of  
$c_{in} - c_{out}$.
{\bf C}  Largest
eigenvalue of $Q$ as a function of $c_{in}-c_{out}$.
The full black line corresponds to $(c_{in}-c_{out}) \frac{q_2}{q_1}$,
where $q_2/q_1 = 8.55$ in the case of this specific
model realization.
{\bf D} Size independent inner products
$\left<v_1|1\right>$ and $\left<v_2|1\right>$  
as  functions
of $c_{in}-c_{out}$. 
}
\label{figS1}
\end{figure*}

\begin{figure*}[!htb]
\includegraphics[height=0.28\textheight]{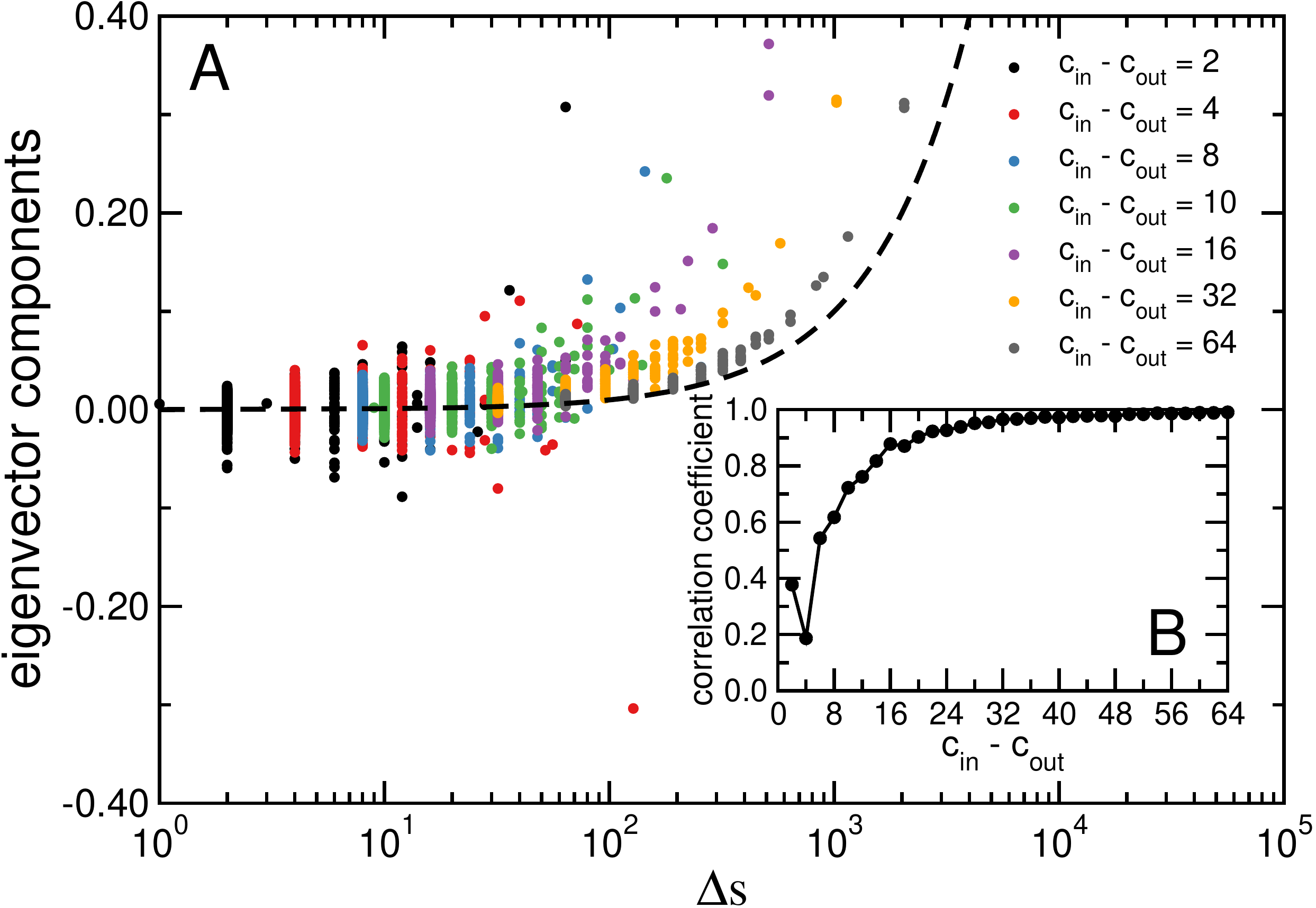}
\qquad
\includegraphics[height=0.28\textheight]{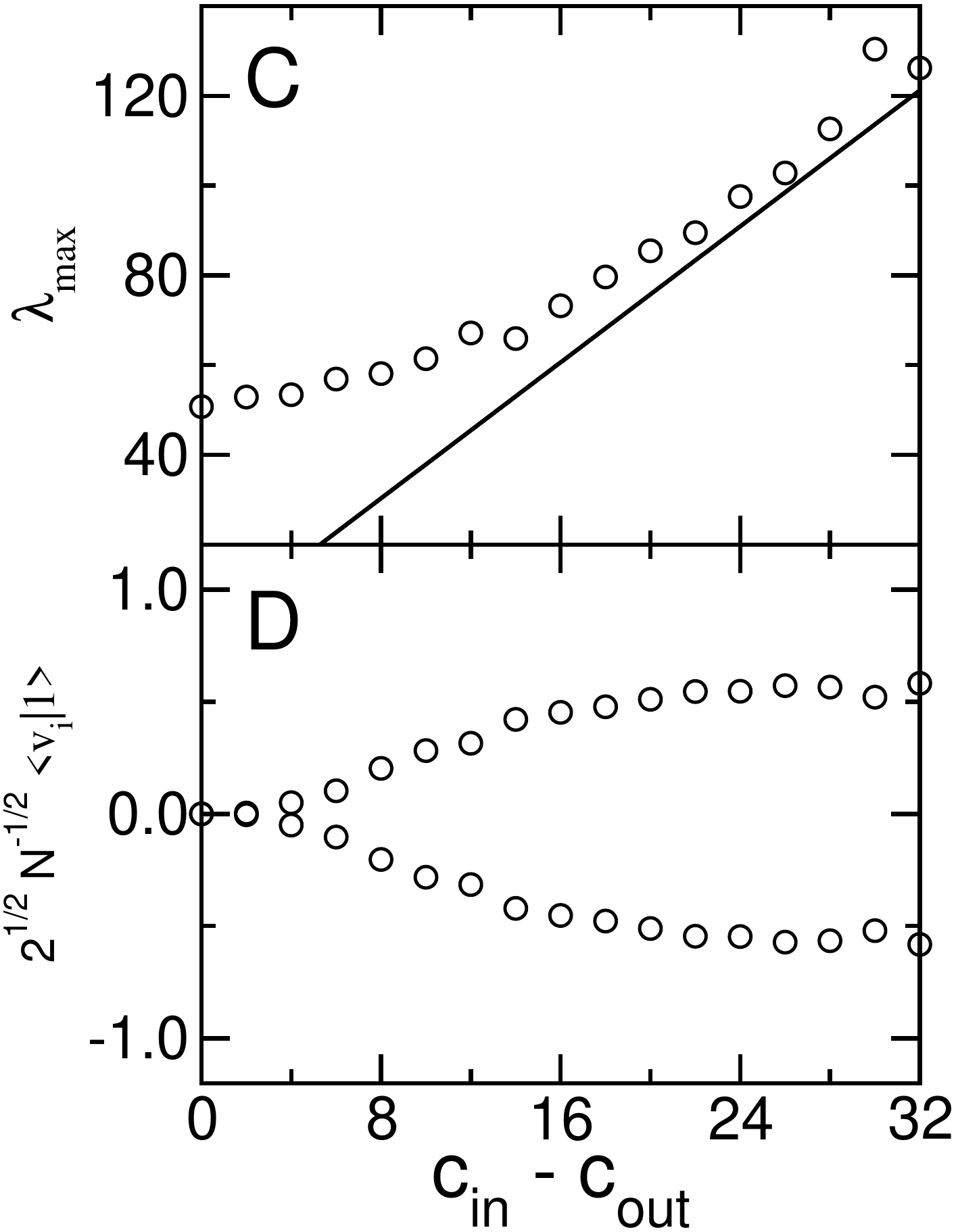}
\caption{Numerical estimation of the 
largest eigenpair of the modularity matrix $Q$
for a single realization of two random scale-free graphs 
with degree exponent $\gamma=2.8$, size $N=1024$,
and internal and external degree parameters such that $c_{in}+c_{out}=64$. 
{\bf A} Components of the principal
eigenvector as functions of the difference between
internal and external degrees. 
We plot here only the components of one of the
two modules, but analogous
results (with opposite sign) 
are valid for the other module.
Different colors corresponds to
different combinations 
of $c_{in}$ and $c_{out}$. The black dashed line
is proportional to $c_{in}-c_{out}$.
{\bf B} Correlation coefficient between
the eigenvector components and the
difference between internal and external degrees 
as a function of  
$c_{in} - c_{out}$.
{\bf C}  Largest
eigenvalue of $Q$ as a function of $c_{in}-c_{out}$.
The full black line corresponds to $(c_{in}-c_{out}) \frac{q_2}{q_1}$,
where $q_2/q_1 = 3.79$ in the case of this specific
model realization.
{\bf D} Size independent inner products
$\left<v_1|1\right>$ and $\left<v_2|1\right>$  
as  functions
of $c_{in}-c_{out}$. 
}
\label{figS2}
\end{figure*}


\end{document}